\documentclass[pre,twocolumn,showpacs,showkeys,a4paper]{revtex4}

\usepackage{graphicx}
\usepackage{amsmath}
\usepackage{bm}

\begin{document}

\title{Response of electrically coupled spiking neurons: a
  cellular automaton approach}
\author{Lucas S. Furtado}
\email{lucas@lftc.ufpe.br}
\affiliation{Laborat\'orio de F\'{\i}sica Te\'orica e Computacional,
  Departamento de F\'{\i}sica, Universidade Federal de Pernambuco,
  50670-901 Recife, PE, Brazil}
\author{Mauro Copelli}
\email{mcopelli@df.ufpe.br}
\thanks{Corresponding author}
\affiliation{Laborat\'orio de F\'{\i}sica Te\'orica e Computacional,
  Departamento de F\'{\i}sica, Universidade Federal de Pernambuco,
  50670-901 Recife, PE, Brazil}

\begin{abstract}
Experimental data suggest that some classes of spiking neurons in the
first layers of sensory systems are electrically coupled via gap
junctions or ephaptic interactions. When the electrical coupling is
removed, the response function (firing rate {\it vs.} stimulus
intensity) of the uncoupled neurons typically shows a decrease in
dynamic range and sensitivity. In order to assess the effect of
electrical coupling in the sensory periphery, we calculate the
response to a Poisson stimulus of a chain of excitable neurons modeled
by $n$-state Greenberg-Hastings cellular automata in two approximation
levels.  The single-site mean field approximation is shown to give
poor results, failing to predict the absorbing state of the lattice,
while the results for the pair approximation are in good agreement
with computer simulations in the whole stimulus range. In particular,
the dynamic range is substantially enlarged due to the propagation of
excitable waves, which suggests a functional role for lateral
electrical coupling. For probabilistic spike propagation the Hill
exponent of the response function is $\alpha=1$, while for
deterministic spike propagation we obtain $\alpha=1/2$, which is close
to the experimental values of the psychophysical Stevens exponents for
odor and light intensities. Our calculations are in qualitative
agreement with experimental response functions of ganglion cells in
the mammalian retina.
\end{abstract}

\pacs{87.18.Sn, 87.19.La, 87.10.+e, 05.45.-a, 05.40.-a}
\keywords{Gap junction, Ephaptic interaction, Olfaction, Retina,
Excitable media, Neural code, Dynamic range, Cellular automata}

\maketitle

\section{Introduction}

Unveiling how neuronal activity represents and processes sensory
information remains a very difficult problem, despite theoretical and
experimental efforts undertaken by neuroscientists for the last
several decades (for a recent review, see~\cite{Kreiman04}). In this
broad context, relatively little attention has been devoted to the
question of how organisms cope with the several orders of magnitude
spanned by the intensities of sensory stimuli~\cite{Cleland99}. This
astonishing ability is most easily revealed in humans by classical
results in psychophysics~\cite{Stevens}: the perception of the
intensity of a given stimulus is experimentally shown to depend on the
stimulus intensity $r$ as $\sim \log(r)$ (Weber-Fechner law) or $\sim
r^{\alpha}$ (Stevens law), where the Stevens exponent $\alpha$ is
typically $<1$.  Those laws have in common the fact that they are {\em
response functions with broad dynamic range\/}, i.e. they map several
decades of stimuli into a single decade of response.

One would like to understand how this broad dynamic range is
physically achieved by neuron assemblies. Recent experimental evidence
suggests that electrical coupling among neurons in the early layers of
sensory systems plays an essential role in weak stimulus
detection. Deans et al.~\cite{Deans02} showed that electrical coupling
is present in the mammalian retina via gap junctions (ionic channels
that connect neighboring cells). In particular, the spiking response
of ganglion cells to light stimulus changes dramatically when the gap
junctions are genetically knocked out: both sensitivity and dynamic
range are reduced~\cite{Deans02}.

Another example comes from the olfactory system. The spiking response
of isolated olfactory sensory neurons (OSNs) to varying odorant
concentration usually presents a narrow dynamic
range~\cite{Rospars00,Rospars03}. This is in contrast with the
response observed in the next layers of the olfactory bulb: both the
glomerular~\cite{Friedrich97,Wachowiak01} and mitral
cell~\cite{Duchamp-Viret90a} responses present a broader dynamic range
than the OSNs. In this case, the tightly packed unmyelinated axons of
OSNs in the olfactory nerve are believed to interact electrically via
ephaptic interactions~\cite{Lowe03} (i.e. mediated by current flow
through the extracellular space), as shown by Bokil et
al.~\cite{Bokil01}.  In particular, their results indicate that a
spike in a single axon can evoke spikes in all other axons of the
bundle, suggesting that some computation is performed prior to the
glomerular layer.

Motivated by these results, previous papers have shown through
numerical simulations that electrical coupling among neurons indeed
changes the response function in a way that is consistent with
experimental results. Due to the coupling, stimuli generate excitable
waves which propagate through the neuron population. The interplay
between wave creation and wave annihilation leads to a nonlinear
amplification of the spiking response, increasing the sensitivity at
low input levels {\em and\/} enhancing the dynamic
range~\cite{Copelli02,Copelli05a}. In one dimension, the robustness of
the mechanism is attested by the diversity of models employed: either
the biophysically realistic Hodgkin-Huxley
equations~\cite{Koch,Copelli02}, a lattice of nonlinear coupled
maps~\cite{Kuva01,Copelli04a,Copelli05a}, or the Greenberg-Hastings
cellular automata (GHCA)~\cite{Copelli02,Copelli05a} yield
qualitatively similar results. The same phenomenon has recently been
observed in simulations with the two-dimensional
GHCA~\cite{Copelli05b}.

In this paper we calculate the response of excitable GHCA model
neurons~\cite{Greenberg78}, where the bidirectional (electrical)
coupling is modeled by a probability $p$ of spike transmission. While
the uncoupled case $p=0$ can be exactly solved, the coupled case $p>
0$ is handled within two mean field approximations, namely at the
single-site and pair levels. The aim is to shed light on the
analytical behavior of the response function for the one-dimensional
case, therefore building on previous efforts which have relied
entirely on numerical simulations.

Our focus on the {\em response\/} of a {\em continuously driven\/}
spatially extended excitable system should be carefully confronted
with other recent studies, where the main interest has been on phase
transitions between an excitable and a self-sustained collective
state. For instance, the SIRS model of epidemics in hypercubic
lattices has been recently investigated under the mean field and pair
approximations~\cite{Joo04b}. In those contagion models, stationary
self-sustained activity becomes stable for sufficiently strong
connection among neighbors, a behavior which has been shown to be
universal under very general assumptions~\cite{Dodds04}. Similar
results have been obtained for a variety of neuronal models, including
collective responses to a localized transient
stimulus~\cite{Hasegawa03a,Roxin04}, as well as the emergence of
sustained activity in complex networks~\cite{Roxin04,Netoff04}.

While interesting in its own, the framework of stable-unstable
collective transitions does not seem particularly suited for our
modeling purposes. To account for sensory responses, the employed GHCA
model is an excitable system which always returns to its absorbing
state in the absence of stimulus, there are no phase transitions. The
refractory period of the GHCA model neurons is absolute (unlike, say,
reaction-diffusion lattices), mimicking the deterministic behavior of
continuous-time systems like the Hodgkin-Huxley equations or
integrate-and-fire models~\cite{Koch}. The only source of
stochasticity of the model regards the firing of the neurons. Stimuli
can come from spiking neighbors (with probability $p$) or from an
``external'' source, which is modeled by a Poisson process and
represents sensory input. Therefore, in the limit $p=1$ the dynamics
is that of a {\em deterministic\/} excitable lattice being {\em
stochastically stimulated\/}, which casts the problem into the
framework of probabilistic cellular automata~\cite{Odor04}.

The paper is organized as follows. In section~\ref{model}, the GHCA
rules are described; section~\ref{uncoupled} contains the exact
calculations for the response of uncoupled neurons, while in
sections~\ref{coupled}~and~\ref{coupledPAIR} results for the coupled
case are discussed in the mean field and pair approximations,
respectively. Our concluding remarks are presented in
section~\ref{conclusion}.

\section{\label{model}The model}

In the $n$-state GHCA model~\cite{Greenberg78} for excitable systems,
the instantaneous membrane potential of the $i$-th cell
($i=1,\ldots,L$) at discrete time $t$ is represented by
$x_i(t)\in\{0,1,\ldots,n-1\}$, $n\geq 3$. The state $x_i(t)=0$ denotes
a neuron at its resting (polarized) potential, $x_i(t)=1$ represents a
spiking (depolarizing) neuron and $x_i(t)=2,\ldots,n-1$ account for
the afterspike refractory period (hyperpolarization). We employ the
simplest rules of the automaton: if $x_i(t)=0$, then $x_i(t+1)=1$ only
if there is a supra-threshold stimulus at site $i$; otherwise,
$x_i(t+1)=0$. If $x_i(t)\geq 1$, then $x_i(t+1)=(x_i(t)+1)\text{mod
}n$, regardless of the stimulus. In other words, the rules state that
a neuron only spikes if stimulated, after which it undergoes an
absolute refractory period before returning to rest.

Whether the neurons are isolated or coupled is implicit in the
definition of the supra-threshold stimulus. We assume {\em external\/}
supra-threshold stimuli to be a Poisson process with rate $r$ (events
per second). Hence at each time step an external stimulus arrives with
probability
\begin{equation}
\label{lambda}
\lambda(r) = 1-e^{-r\tau}
\end{equation}
per neuron. Notice that $\tau=1$~ms corresponds to the approximate
duration of a spike and is the time scale adopted for the time step of
the model. The number of states $n$ therefore controls the duration of
the refractory period (which corresponds to $n-2$, in ms). In the
biological context, $r$ could be related for example with the
concentration of a given odorant presented to an olfactory
epithelium~\cite{Rospars00}, or the light intensity stimulating a
retina~\cite{Deans02}. We shall refer to $r$ as the stimulus rate or
intensity.

When electrically coupled, neurons at rest can also be stimulated by
their neighbors. We define $p$ and $q$ as the probabilities that a
resting neuron spikes as a consequence of transmission (ionic current
flow) from respectively one or two spiking neighbors [see
Eq.~(\ref{eq:p1coupled})]. We keep $p$ and $q$ as two independent
parameters in most calculations to show the robustness of some
asymptotic results. In the simulations, we concentrate on the more
physically intuitive choice of $q = 1 - (1-p)^2$, where the
contributions from two spiking neighbors are independent.

Let $P_{t}^{(i)}(k)$ be the probability that the $i$-th neuron is in
state $k$ at time $t$. Since the dynamics of the refractory state is
deterministic, the equations for $k\geq 2$ are simply

\begin{eqnarray}
\label{eq:refrac}
P_{t+1}^{(i)}(2) & = & P_{t}^{(i)}(1) \nonumber \\
P_{t+1}^{(i)}(3) & = & P_{t}^{(i)}(2) \nonumber \\
\ & \vdots & \ \nonumber \\	
P_{t+1}^{(i)}(n-1) & = & P_{t}^{(i)}(n-2) \; .
\end{eqnarray}
To describe the coupling among first neighbors, let
$P_{t}^{(i)}(k,l,m)$ be the joint probability that sites $i-1$, $i$
and $i+1$ are respectively in the states $k$, $l$ and $m$ at time
$t$. Following the definitions of $\lambda$, $p$ and $q$ above, the
equation for $P_{t}^{(i)}(1)$ thus becomes

\begin{eqnarray}
\label{eq:p1coupled}
P_{t+1}^{(i)}(1) & = & [1-(1-\lambda)(1-q)]P_{t}^{(i)}(1,0,1) \nonumber \\ 
& & + [1-(1-\lambda)(1-p)]\left(\sum_{k\neq 1}^{n-1}P_{t}^{(i)}(1,0,k)
\right. \nonumber \\
&& \left. + \sum_{k\neq 1}^{n-1} P_{t}^{(i)}(k,0,1)\right) \nonumber \\ 
& & + \lambda\sum_{k\neq1}^{n-1}\sum_{l\neq 1}^{n-1} P_{t}^{(i)}(k,0,l) \; .
\end{eqnarray}
Finally, the dynamics for $P_{t}^{(i)}(0)$ can be obtained by the
normalization condition
\begin{equation}
\label{eq:norm}
\sum_{k=0}^{n-1}P_{t}^{(i)}(k) = 1 \; ,\forall t,i\; ,
\end{equation}
which completes the set of equations for one-site probabilities.

It is reasonable to assume homogeneity in the system when
$L\to\infty$, so we can drop the superscript $(i)$ in
Eqs.~(\ref{eq:refrac}-\ref{eq:norm}) and in what follows. We also
expect isotropy (right-left symmetry) in the probabilities:
$P_{t}(k,l) = P_{t}(l,k)$, $P_{t}(k,l,m)=P_{t}(m,l,k)$ etc. Recalling
the normalization condition
$\sum_{j_1=0}^{n-1}P_{t}(j_1,j_2,\ldots,j_m) = P_{t}(j_2,\ldots,j_m)$,
one can rewrite Eq.~(\ref{eq:p1coupled}) as

\begin{eqnarray}
\label{eq:p1}
P_{t+1}(1) & = & \lambda P_{t}(0) + 2p(1-\lambda)P_{t}(1,0)  \nonumber\\ 
& &+ (1-\lambda)(q-2p)P_{t}(1,0,1)\; .
\end{eqnarray}

The stationary value of any joint probability will be denoted by
omitting the subscript $t$, thus $P(\bullet)\equiv \lim_{t\to\infty}
P_{t}(\bullet)$. We start by solving Eqs.~(\ref{eq:refrac})
and~(\ref{eq:norm}) in the stationary state, which together yield

\begin{equation}
\label{eq:p1p0}
P(0) = 1 - (n-1)P(1)\; ,
\end{equation}
a result which is exact and holds $\forall p,q$.

We are interested in obtaining the behavior of $P(1)$ as a function of
$\lambda$ (or $r$). Note that $P(1)$ coincides with the average firing
rate per neuron (measured in spikes per ms, according to the choice of
$\tau$) in the limit $L,t\to\infty$. In simulations, firing rates have
been calculated by division of the total number of spikes in the chain
by $LT$, where $T \sim {\cal O}(10^5)$ and $L \sim {\cal O}(10^5)$
were the typical number of time steps and model neurons
employed~\cite{Copelli02}. We define $F(\lambda)\equiv P(1)$ as the
response function of the system.

Due to the absolute nature of the refractory period, the maximum
firing rate of the model neurons is $F_{max}\equiv 1/n$, a result
which is easily obtained $\forall p,q$ by setting $\lambda=1$ in
Eqs.~(\ref{eq:p1})~and~(\ref{eq:p1p0}). The dynamic range
$\delta_{\lambda}$ of the response curve $F(\lambda)$ follows the
definition commonly employed in biology~\cite{Firestein93,Rospars00}:
\begin{equation}
\label{eq:dr}
\delta_{\lambda} = 10\log_{10}\left(\frac{\lambda_{0.9}}{\lambda_{0.1}}\right)\; ,
\end{equation}
where $\lambda_{x}$ satisfies 
\begin{equation}
\label{eq:lambdmax}
F(\lambda_{x})=xF_{max}\; .
\end{equation}
The dynamic range is therefore the number of decibels of input which
are mapped into the $\simeq 9.5~\text{dB}$ of output comprised in the
$[0.1F_{max},0.9F_{max}]$ interval (see
Fig.~\ref{fig:respostalinlog}). In the biological context of the
model, it measures the ability of the system to discriminate different
orders of magnitude of stimulus intensity. We will show below that if
one chooses to calculate $\delta_{r}$ using $r_{x}\equiv
-\tau^{-1}\ln(1 - \lambda_{x})$ instead of $\lambda_{x}$ in
Eq.~\ref{eq:dr}, results are essentially unchanged.

\section{\label{uncoupled}Uncoupled neurons}

\begin{figure}
\includegraphics[width=0.8\columnwidth]{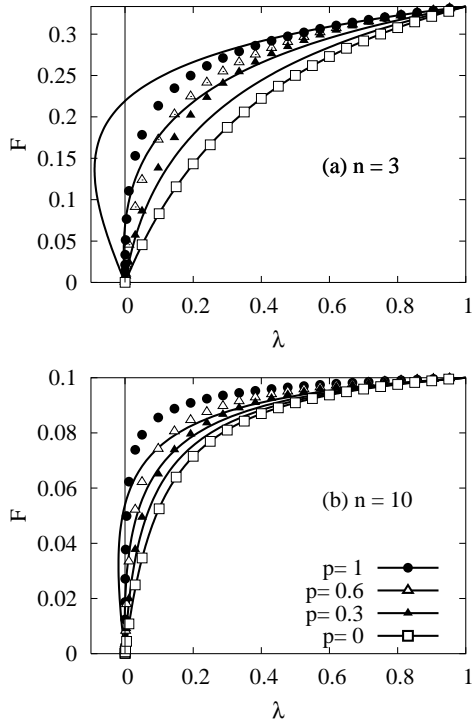}
\caption{\label{fig:campomedio}Response curves for (a) $n=3$ and (b)
$n=10$ automata: simulations (symbols) and mean field approximation
[lines, according to Eq.~(\ref{eq:p1ss})]. From bottom to top, $p=0$,
$0.3$, $0.6$ and $1$, $q=1-(1-p)^2$.  In the simulations, standard
deviations over 10 runs are smaller than symbol sizes, so error bars
are omitted in all figures.  Notice the negative slope and
multi-valuedness of the single-site approximation for $p>1/2$ and
$\lambda\leq 0$.}
\end{figure}

\begin{figure}
\includegraphics[width=0.8\columnwidth]{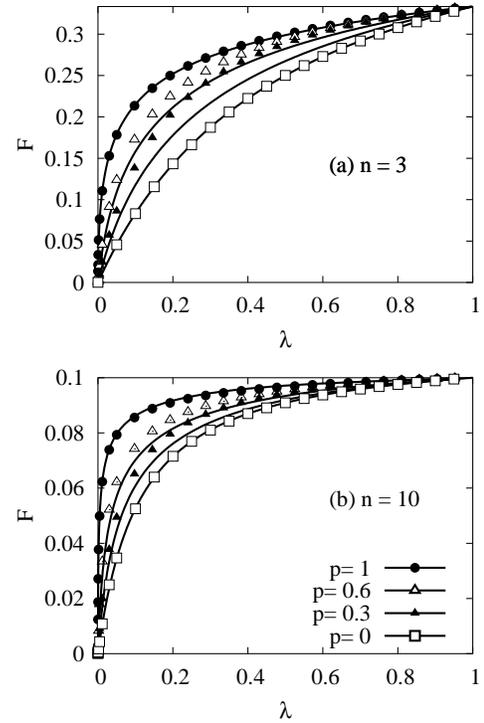}
\caption{\label{fig:pares}Response curves for (a) $n=3$ and (b) $n=10$
automata: simulations (symbols) and pair approximation [lines,
according to Eqs.~(\ref{eq:2}-\ref{eq:1})]. From bottom to top, $p=0$,
$0.3$, $0.6$ and $1$, $q=1-(1-p)^2$. The pair approximation eliminates
the small-$\lambda$ anomalies of the single-site solution, yielding
excellent agreement with simulations for the extreme cases $p=0$ and
$p=1$.}
\end{figure}

The uncoupled case $p=q=0$ can be exactly solved by taking the limit
$t\to\infty$ in Eq.~(\ref{eq:p1}) which, together with
Eq.~(\ref{eq:p1p0}), yields

\begin{equation}
\label{eq:linsat}
P(1) = f(\lambda) = \frac{\lambda}{1 + (n-1)\lambda}\; .
\end{equation}
This linear saturating response is depicted for $n=3$
(Figs.~\ref{fig:campomedio}-\ref{fig:respostalinlog}) and $n=10$
(Figs.~\ref{fig:campomedio}-\ref{fig:pares}), in complete agreement
with simulations. It belongs to the family of Hill functions defined
by $H_{\alpha}(x)\equiv Cx^\alpha/(x_0^\alpha+x^\alpha)$, where the
Hill exponent in this case is $\alpha=1$.

The dynamic range can be promptly calculated: $\delta_{\lambda}(n) =
10\log_{10} \left\{ \left[ 1+9n \right] / \left[ 1+n/9 \right]
\right\}$ and $\delta_{r}(n) = 10\log_{10} \left\{ \ln\left[ 1+9/n
\right] / \ln\left[ 1+1/(9n) \right] \right\}$, both of which rapidly
converge to $10\log_{10}(81)\simeq 19$~dB for moderate values of $n$
(see lower curves in Fig.~\ref{fig:faixadinamica}). As we shall see,
the electrical coupling can lead to dynamic ranges typically twice as
large.

\section{\label{coupled}Coupled Neurons: Mean Field Approximation}

As can be seen in Eq.~(\ref{eq:p1}), $P_t(1)$ depends on two- and
three-site probabilities, and in general $k$-site probabilities depend
on up to $(k+2)$-site probabilities. The dynamical description of the
system thus requires an infinite hierarchy of equations. The mean
field approximation at the single-site level corresponds to the
simplest truncation of this hierarchy, and consists in discarding the
influence of all neighbors in the conditional
probabilities~\cite{Atman03}, thus $P_{t}(j_1|j_2,\ldots,j_m) \approx
P_{t}(j_1)$, which leads to
\begin{equation}
\label{eq:singlesite}
P_{t}(j_1,\ldots,j_m) \approx \prod_{k=1}^{m}P_{t}(j_k) \; .
\end{equation}
In this approximation, Eq.~(\ref{eq:p1}) becomes
\begin{eqnarray}
\label{eq:p1ss}
P_{t+1}(1) & \approx & P_{t}(0)\left\{ \lambda + 2p(1-\lambda)P_{t}(1)
 \right. \nonumber \\
&& \left. +(q-2p)(1-\lambda)P_{t}(1)^2 \right\}\; ,
\end{eqnarray}
which, together with Eq.~(\ref{eq:p1p0}), can be used to eliminate
$P(0)$ and render $P(1)=F(\lambda)$ implicitly through the relation

\begin{equation}
\label{eq:lambf}
\lambda \approx \frac{(1-2p)F + (2pn-q)F^2 + 
(n-1)(q-2p)F^3}{[1-(n-1)F][1-2pF+(2p-q)F^2]} \; .
\end{equation}
As a consistency check, notice that setting $p=q=0$ in
Eq.~(\ref{eq:lambf}) recovers Eq.~(\ref{eq:linsat}) (in other words,
mean field is exact for the uncoupled case, as it should). However,
for $0<p,q\leq 1$, $F(\lambda)$ as given by Eq.~(\ref{eq:lambf})
yields in general a poor agreement with numerical simulations, as can
be seen in Fig.~\ref{fig:campomedio} for different values of $p$. When
$\lambda \simeq 0$, Eq.~(\ref{eq:lambf}) predicts $F\simeq
\lambda/(1-2p)$, which leads to obviously nonphysical results for
$p\geq 1/2$ (see leftmost part of Fig.~\ref{fig:campomedio}). In
particular, $F(\lambda)$ is multi-valued, leading to $\lim_{\lambda\to
0^{+}} F \neq 0$. The mean field result therefore suggests a
transition to an ordered state at $\lambda=0$ which is simply
forbidden by the automaton rules~\cite{lewis00}. By generalizing
Eq.~(\ref{eq:p1ss}), this failure to predict the absorbing state of
the system can in fact be extended to regular lattices with
coordination $z$, where the single-site approximation yields
$F\stackrel{\lambda\to 0}{\simeq} \lambda/(1-pz)$. Since this level of
approximation is clearly not satisfactory for the calculation of the
dynamic range, a refinement is needed.

\section{\label{coupledPAIR}Coupled Neurons: Pair Approximation}

The pair approximation consists in keeping the influence of only one
neighbor in the conditional probabilities~\cite{Atman03}, thus
$P_{t}(j_1|j_2,\ldots,j_m) \approx P_{t}(j_1|j_2)$. In this case
$m$-site probabilities are reduced to combinations of up to two-site
probabilities. In particular, three- and four-site probabilities
become~\cite{Atman03}
\begin{subequations}
\label{eq:pair}
\begin{eqnarray}
P(k,l,m) & \approx & \frac{P(k,l)P(l,m)}{P(l)} \\
P(j,k,l,m) & \approx & \frac{P(j,k)P(k,l)P(l,m)}{P(k)P(l)} \; .
\end{eqnarray}
\end{subequations}
It is therefore possible to rewrite Eq.~(\ref{eq:p1}) in this
approximation: 

\begin{equation}
\label{eq:p1pair}
P_{t+1}(1) \approx \lambda P_{t}(0) + (1-\lambda)P_{t}(1,0)\left[2p
+ (q-2p)\frac{P_{t}(1,0)}{P_{t}(0)}\right]\; .
\end{equation}
Eq.~(\ref{eq:p1pair}), on its turn, depends on $P_{t}(1,0)$, whose
evolution can be exactly obtained (up to homogeneity and isotropy
assumptions):

\begin{eqnarray}
\label{eq:p10}
P_{t+1}(1,0) & = & \lambda P_{t}(n-1,0) + p(1-\lambda)P_{t}(n-1,0,1)
\nonumber \\
& & + \lambda(1-\lambda)P_{t}(0,0) \nonumber \\
&& + p(1-\lambda)(1-2\lambda)P_{t}(1,0,0) \nonumber \\
&&  - p^2(1-\lambda)^2 P_{t}(1,0,0,1) \; . \nonumber \\
\end{eqnarray}
With the help of the pair approximation in Eqs.~(\ref{eq:pair}),
Eq.~(\ref{eq:p10}) becomes

\begin{eqnarray}
\label{eq:p10pair}
P_{t+1}(1,0) & \approx & P_{t}(n-1,0) \left[\lambda +  
  p(1-\lambda)\frac{P_{t}(1,0)}{P_{t}(0)}\right] \nonumber \\
&& +  (1-\lambda)P_{t}(0,0)\left[\lambda + 
p(1-2\lambda)\frac{P_{t}(1,0)}{P_{t}(0)} \right. \nonumber \\
&& \left. -  p^2(1-\lambda)\frac{P_{t}(1,0)^2 }{P_{t}(0)^2} \right]\; .
\end{eqnarray}
Since $P_{t}(j,0)$ depends on $P_{t}(j-1,0)$ and $P_{t}(j-1,n-1)$, and
$P_{t}(0,0)$ depends, among others, on $P_{t}(n-1,n-1)$, all the
equations for two-site probabilities are in principle required for the
dynamical description of the system. Together with the equations for
single-site probabilities, they form a $(n^2+3n)/2$-dimensional map
whose stationary stable solution can be analytically studied. While
the Appendix contains details of the derivation of those equations, we
discuss the main results below.

The main point to be noted is that the calculation of the stationary
state presents additional difficulties when $n\geq 4$. In that case,
the pair probabilities $P(j,0)$ with $2\leq j \leq n-2$ have the same
stationary value, but differ from $P(n-1,0)$. In particular, for
$p=q=1$ one obtains $P(j,0)=0$ [$2\leq j \leq n-2$, see
Eq.~(\ref{eq:J})], which in turn leads to many other vanishing
probabilities and gives the deterministic case a sparse stationary
matrix [see
Eqs.~(\ref{eq:p21}),~(\ref{eq:pj1})~and~(\ref{eq:p1j})]. Those terms
do not exist for the $n=3$ case, which makes its analysis considerably
simpler. In either case, for $n\geq 3$ one obtains the reasonable
result $P(n-1,0) \approx P(1,0)$, the l.h.s. (r.h.s.)  being
associated to the end (beginning) of an excitable wave front [see
Eq.~(\ref{eq:pdelta})]. Combining these results, a normalization
condition and the linearity of Eq.~(\ref{eq:p10pair}) in $P_{t}(0,0)$,
we obtain (see Appendix):

\begin{widetext}
\begin{eqnarray}
\label{eq:2}
&& P(0) - P(1,0)\left\{2 + (n-3)\left[\frac{(1-p)P(0)+(p-q)P(1,0)}{P(0)-pP(1,0)}\right] \right\}
    \nonumber \\ 
&\approx & \frac{P(1,0)P(0)[P(0)-pP(1,0)]}{\lambda P(0)^2 +
  p(1-2\lambda)P(0)P(1,0) - p^2(1-\lambda)P(1,0)^2}\; ,
\end{eqnarray}
which is valid $\forall n\geq 3$.  Consider now the stationary state
of Eqs.~(\ref{eq:p1p0})~and~(\ref{eq:p1pair}). They can be combined in
a quadratic equation for $P(1,0)$, yielding
\begin{eqnarray}
\label{eq:1}
&& (2p-q) P(1,0) \approx   G_{\pm}(P(0)) \nonumber \\
&& \equiv pP(0) \pm \sqrt{\frac{P(0)\left\{P(0)[(n-1)p^2+2p-q + 
\lambda(n-1)(2p-p^2-q)]+(q-2p)\right\}}{(n-1)(1-\lambda)}}\; .
\end{eqnarray}
\end{widetext}
Since $P(1,0)$ must vanish $\forall p,q$ in the limit $\lambda\to 0$,
$G_{-}$ is the only acceptable solution.

\begin{figure}[h]
\includegraphics[width=0.99\columnwidth]{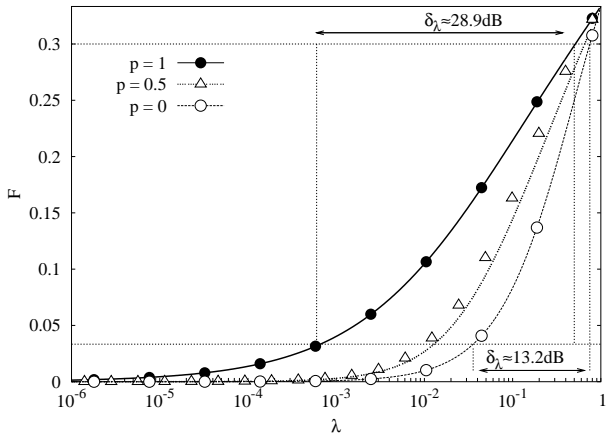}
\caption{\label{fig:respostalinlog}Linear-log plot of the response
curve for $n=3$ automata with $p=q=1$ (filled circles), $p=0.5$ and
$q=1-(1-p)^2$ (open triangles), and $p=q=0$ (open circles). Lines
correspond to the pair approximation. Horizontal lines are
$F=0.1F_{max}$ and $F=0.9F_{max}$, vertical lines are
$\lambda=\lambda_{0.1}$ and $\lambda=\lambda_{0.9}$ and arrows
illustrate the dynamic range $\delta_{\lambda}$ [Eq.~(\ref{eq:dr})]
for $p=0$ and $p=1$. The dynamic range of a chain of neurons with
deterministic spike propagation is about twice as large as that of its
uncoupled counterpart.}
\end{figure}

\begin{figure}[h]
\includegraphics[width=0.99\columnwidth]{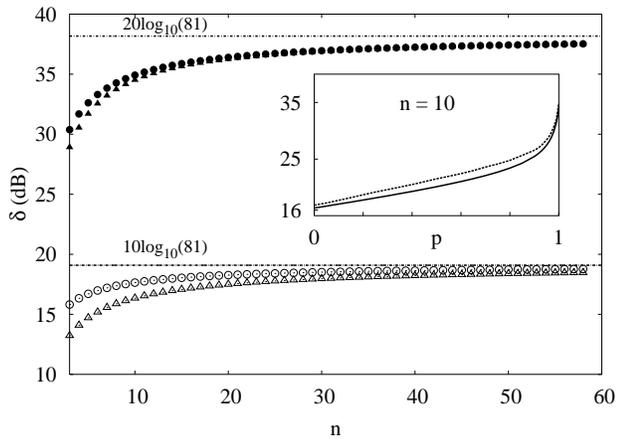}
\caption{\label{fig:faixadinamica}Dynamic ranges (triangles for
$\delta_\lambda$, circles for $\delta_r$) as a function of the number
of states of the GHCA, obtained from the stationary solution of the
pair approximation. Open (filled) symbols correspond to the $p=q=0$
($p=q=1$) case. Inset: $\delta_\lambda$ as a function of $p$ for
$n=10$ for simulations (dashed line) and pair approximation (solid
line). In spite of the underestimation of the response observed in
Fig.~\ref{fig:pares}, the pair approximation is able to reproduce the
behavior of the dynamic range as a function of the probability of
spike transmission. }
\end{figure}

\begin{figure}[h]
\includegraphics[width=0.99\columnwidth]{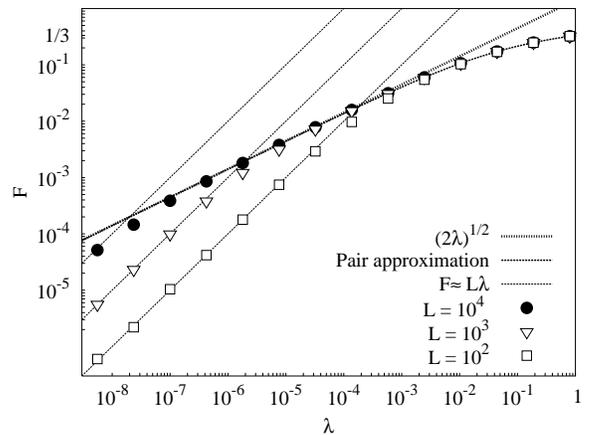}
\caption{\label{fig:respostaloglog}Log-log plot of the response curve
for $p=q=1$. Pair approximation (solid lines) and simulations
(symbols) follow a power law ($\alpha=1/2$) for weak stimuli, while
finite size effects lead to a linear response $F\simeq L\lambda$
(dotted lines) for $\lambda\lesssim\lambda_c(L)$.}
\end{figure}

The solution of Eqs.~(\ref{eq:2})~and~(\ref{eq:1}) determines $P(0)$
as a function of $\lambda$.  Instead of numerically solving them, we
iterate the $(n^2+3n)/2$-dimensional map involving the one- and
two-site probabilities for each value of $\lambda$ until it converges
to its stationary state. Despite the growing number of equations with
$n$, this method has the advantage of avoiding unstable fixed
points~\cite{Atman03} [Eqs.~(\ref{eq:2})~and~(\ref{eq:1}) can have
more than one solution]. Once $P(0)$ is known, the response $P(1) =
F(\lambda)$ is obtained via Eq.~(\ref{eq:p1p0}).

\subsection{\label{p=1}Deterministic spike propagation ($p=1$)}

Ordinary differential equations (ODEs) are the standard modeling tool
in computational neuroscience. This is due to the fact that, despite
the stochastic nature of the opening and closing of individual ionic
channels, a neuron containing a large number of such channels can very
often be extremely well described by a deterministic
dynamics~\cite{Koch} (an approach which has been established since the
seminal work of Hodgkin and Huxley~\cite{HH52}). In the present
context, it is therefore important to address the case $p=1$. This
limit is consistent with a variety of scenarios in which, in addition
to the dynamics of individual neurons, spike transmission is also well
described by deterministic behavior. Specifically regarding our
present study, deterministic spike transmission due to electrical
coupling has previously been employed in the literature to model
axo-axonal interactions both via ephaptic interactions (e.g. in the
olfactory nerve~\cite{Bokil01}) and gap junctions (e.g. in the
hippocampus~\cite{Traub99b,lewis00}). This is in contrast with, say,
dendro-dendritic gap junctions or chemical synapses (in the latter
case, synaptic transmission can sometimes be as low as 10\% due to the
inherent stochasticity in the process of neurotransmitter
release~\cite{Kandel,Koch}), where the $p=1$ limit can hardly be
expected to apply. As we shall see in the following, in addition to
its biological relevance, the response function for $p=1$ also has a
different characteristic exponent which will help us understand the
limiting behavior for $p\lesssim 1$.

Figure~\ref{fig:pares} shows the excellent agreement between the pair
approximation and the simulations when $p=q=1$. One observes that the
response is particularly enhanced in the low stimulus range. This
feature is best seen in the logarithmic scale of
Fig.~\ref{fig:respostalinlog}: in comparison with the uncoupled case
$p=0$, the effect of the electrical interaction is to increase the
sensitivity of the response for more than a decade, leading to a
dramatic rise of the dynamic range.

For each value of $n$, we can thus obtain the stationary response
$F(\lambda)$ and the dynamic ranges $\delta_\lambda$ and $\delta_r$ in
the pair approximation. Even though the response curve changes
considerably for varying $n$ (since $F$ is bounded by $F_{max} = 1/n$,
see Fig.~\ref{fig:pares}), the dynamic range levels off smoothly, as
can be seen in Fig.~\ref{fig:faixadinamica}. For increasing $n$, the
dynamic range of the $p=q=1$ case approaches twice the value for the
uncoupled case. The fact that this result holds for both $\delta_r$
and $\delta_\lambda$ can be understood on the basis of the
low-stimulus amplification, which plays the central role in the
phenomenon: in this regime $\lambda$ is approximately linear in
$r$. Should one choose a different relationship $\lambda(r)$,
$\delta_r$ would obviously have different values, but the drastic
enhancement in the response due to the electrical coupling would not
be affected.


In order to understand the low-stimulus amplification induced by the
coupling, we have analyzed Eqs.~(\ref{eq:2}-\ref{eq:1}) when $\lambda
\simeq 0$. Inspection of Fig.~\ref{fig:pares} and previous numerical
simulations~\cite{Copelli02} suggest that $P(1)\simeq
C\lambda^\alpha$, with $\alpha<1$. This ansatz can be inserted into
Eqs.~(\ref{eq:p1p0})~and~(\ref{eq:2}-\ref{eq:1}) for general $p$ and
$q$, yielding $\alpha=1/2$ {\em and\/} $p=1$ as
solutions. Deterministic spike propagation therefore leads to a power
law response
\begin{equation}
\label{eq:sqrt}
F(\lambda) \stackrel{\lambda\to 0}{\simeq} \sqrt{2\lambda}\; ,
\end{equation}
a result that holds $\forall n,q$, as should be expected.  This power
law suggests a Hill function with $\alpha=1/2$, which is an excellent
approximation for $F(\lambda)$ in the whole $\lambda$ interval when
$n$ is large. This result explains the doubling of the dynamic range
as compared to the uncoupled case and is reminiscent of
reaction-diffusion processes modeled by lattice
gases~\cite{Stinchcombe93,Grynberg94,Oliveira99,Mendonca98} and
partial differential equations~\cite{Ohta05}. Since the Hill function
can be regarded as a saturating Stevens law, it is interesting to note
that the experimental values of the Stevens exponents for light and
smell intensities are respectively $\alpha\simeq 0.5$ and
$\alpha\simeq 0.6$~\cite{Stevens}.

Let us now consider a chain with {\em finite\/} $L$ and a very small
value of $\lambda$ such that a single external stimulus occurs in a
given time interval. In this case, the deterministic nature of the
propagation would lead to $L$ spikes in the chain, while a single
spike would be observed if the neurons were uncoupled. One would thus
have $F\simeq Lf$, and since $f\stackrel{\lambda\to 0}{\simeq}\lambda$
[from Eq.~(\ref{eq:linsat})] we obtain $F\stackrel{\lambda\to
0}{\simeq}L\lambda$. This corresponds to a linear regime where
excitable waves do not interact. If one increases $\lambda$, waves
will start annihilating each other, leading to the power law response
of Eq.~(\ref{eq:sqrt}), as can be clearly seen in
Fig.~\ref{fig:respostaloglog}. For a given system size $L$, there is
therefore a crossover value $\lambda_c(L)\simeq 2/L^2$ from a linear
to a nonlinear response. In an infinite chain, there is no linear
response since for any nonzero stimulus rate two excitable waves will
inevitably interact.

\begin{figure}[t]
\includegraphics[width=0.99\columnwidth]{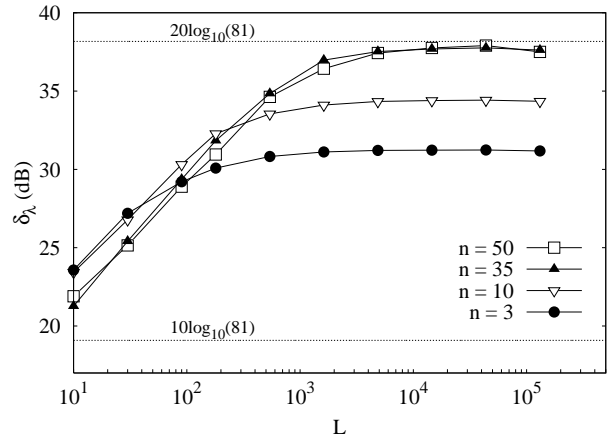}
\caption{\label{fig:faixadinamicaL}Dynamic range as a function of the
system size $L$ for $p=q=1$. Lines are just guides to the eye.}
\end{figure}

To assess the finite size effects in the biological context of the
model, we notice that the dynamic range will be affected only if
$\lambda_c(L) \gtrsim \lambda_{0.1}$, that is, for $L\lesssim 20n$.
For neurons with refractory periods of the order of tens of ms,
neuronal assemblies with $L \gtrsim 10^{3-4}$ should therefore be well
approximated by the limit $L\to\infty$, as can be seen in
Fig.~\ref{fig:faixadinamicaL}. It is important to emphasize, however,
that even small chains dominated by finite size effects still possess
dynamic ranges which are significantly larger than those of the
uncoupled case. For $\lambda_{0.1} \lesssim \lambda_c(L)$, the dynamic
range increases approximately logarithmically with the total number of
connected neurons, a result which holds for regular lattices in any
dimension~\cite{Copelli05b}.

\subsection{\label{pneq1}Probabilistic spike propagation ($p\neq 1$)}

For $p\neq 1$, communication between spiking and resting neurons may
eventually fail. This provides us with the simplest test under which
the robustness of the mechanism for dynamic range enhancement can be
checked. From the biological point of view, this regime could be
useful for modeling networks of neurons connected by chemical
synapses, for instance.

We start the analysis of the $p\neq 1$ case by noticing in
Figs.~\ref{fig:pares}~and~\ref{fig:respostalinlog} that the agreement
between simulations and the pair approximation is better than the mean
field results (specially in the low-stimulus region), but certainly
not so good as in the extreme cases $p=0$ and $p=1$. This inevitably
affects the estimation of the dynamic range via the stationary state
of the pair approximation (see below), but nonetheless allows us to
understand qualitatively how the response changes as $p$ varies.

As pointed out in the preceding section, the dynamic range is enhanced
for $p=1$ primarily due to the low-stimulus amplification associated
to the propagation of excitable waves. As opposed to the deterministic
case, however, for $p\neq 1$ a single excitable wave traveling in an
infinite chain initially at rest will eventually die out. We should
therefore expect a qualitative change in the response function for
$\lambda\simeq 0$.  This is indeed confirmed by reinserting the ansatz
$P(1)\simeq C\lambda^\alpha$ in
Eqs.~(\ref{eq:p1p0})~and~(\ref{eq:2}-\ref{eq:1}) without the
constraint $\alpha<1$. In this case, the linear behavior suggested by
the plots in Fig.~\ref{fig:pares} is easily confirmed:

\begin{figure}[b]
\includegraphics[width=0.99\columnwidth]{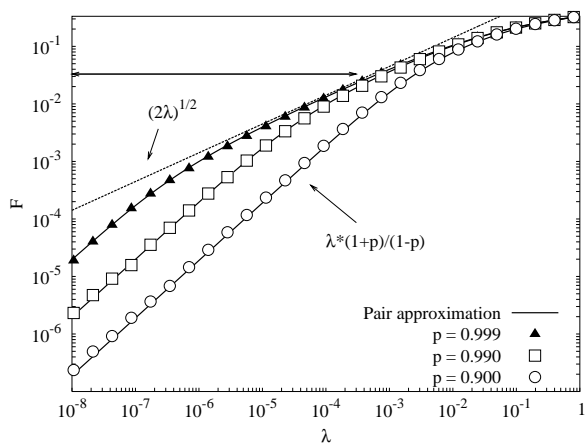}
\caption{\label{fig:respostalogpneq1}Log-log plot of the response
curve: pair approximation (solid lines) and simulations (symbols) with
$q=1-(1-p)^2$ and $n=3$. For $p\lesssim 1$, there is a crossover
between $\alpha=1$ and $\alpha=1/2$. The horizontal arrow shows
$0.1F_{max}$.}
\end{figure}

\begin{equation} 
\label{eq:1+p}
F(\lambda) \stackrel{\lambda\to 0}{\simeq}
\left(\frac{1+p}{1-p}\right)\lambda \; ,
\end{equation}
which is again valid $\forall n,q$. Therefore, the low-stimulus
response for $p<1$ is governed by $\alpha=1$, which is confirmed by
the simulations displayed in
Fig~\ref{fig:respostalogpneq1}. Interestingly, such a change in
exponent for $p<1$ seems to be absent from reaction-diffusion models
in lattice gases~\cite{Stinchcombe93,Grynberg94,Oliveira99,Mendonca98}
as well as partial differential equations~\cite{Ohta05}.

Thanks to the growing coefficient in Eq.~(\ref{eq:1+p}), for
$p\lesssim 1$ the proximity to the transition that occurs at $p=1$
produces a crossover in the response from a linear to a square root
behavior, dismissing the suspicion that a larger exponent might
severely deteriorate the enhancement of the dynamic range (see
Fig.~\ref{fig:respostalogpneq1}).  In particular, notice that, for
$p\lesssim 1$, $\alpha=1/2$ is the dominant exponent at
$F=0.1F_{max}$, which is used to calculate the dynamic range (see
horizontal arrow in Fig.~\ref{fig:respostalogpneq1}). This explains
the smooth monotonic increase in $\delta_\lambda$ with $p$, as shown
in the inset of Fig.~\ref{fig:faixadinamica}, even though the exponent
changes discontinuously at $p=1$. On the one hand, we observe that
deterministic spike propagation ($p=1$) is certainly not essential for
the enhancement of the dynamic range, in the sense that any $p>0$
yields a better response than uncoupled neurons. On the other hand, it
is interesting to point out that, as $p$ is varied from 0 to 1, the
increase in dynamic range is particularly pronounced for $p\gtrsim
0.9$.  This is in agreement with the conjecture that the reliability
of electrical coupling among spiking neurons could indeed play a
significant role in early sensory processing.

\section{\label{conclusion}Concluding remarks}

We have calculated the collective response to a Poisson stimulus of a
chain of electrically coupled excitable neurons modeled by $n$-state
Greenberg-Hastings cellular automata. The single-site mean field
approximation has been shown to give poor results, failing to predict
the absorbing state of the lattice in the absence of stimulus for
$p\geq 1/2$. The pair approximation yields a response curve which
agrees reasonably well with simulations {\em in the whole stimulus
range\/}. It is interesting to remark that the agreement is
particularly good when $p=q=1$, a deterministic regime in which the
GHCA lattice mimics a system of coupled nonlinear ODEs. This
reinforces an interesting perspective in the context of computational
neuroscience: the possibility of applying techniques from
nonequilibrium statistical mechanics to the study of spatially
extended nonlinear systems.

The enhancement of the dynamic range in the presence of electrical
coupling is due to low-stimulus amplification. For uncoupled neurons
($p=0$) the response is governed by the Hill exponent $\alpha=1$,
leading to a dynamic range of $\sim 19$dB. For coupled neurons this
value can be doubled in the limit $p=q\to 1$, when the Hill exponent
becomes $\alpha=1/2$. This value is close to Stevens exponents
observed in psychophysical experiments of smell and light intensities.
For $0<p<1$, the exponent remains $\alpha=1$, but the dynamic range
increases smoothly, which can be understood on the basis of the
crossover behavior observed in the response function for $p\lesssim
1$. 

In the context of experiments at the cellular level, the enhancement
of the dynamic range associated with an increase in sensitivity is
also observed in both the olfactory~\cite{Wachowiak01} and
visual~\cite{Deans02} systems. While the dynamic range of OSNs (the
neurons which perform the initial transduction) is about $\sim
10$dB~\cite{Rospars00,Rospars03}, the glomeruli (the next processing
layer) have dynamic ranges at least twice as
large~\cite{Wachowiak01}. It remains to be investigated experimentally
whether this enhancement is indeed due to ephaptic interactions among
the unmyelinated OSN axons in the olfactory nerve.

\begin{figure}[b]
\includegraphics[width=0.99\columnwidth]{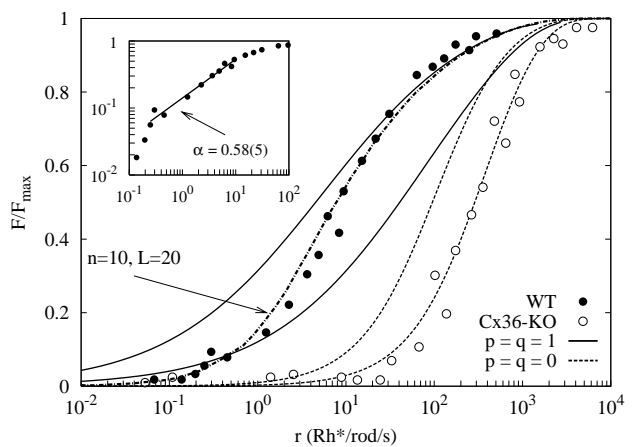}
\caption{\label{fig:deans}Experimental response curves (normalized
firing rate vs. light intensity) of retinal on-center ganglion cells
in linear-log (main plot) and log-log (inset) scales (data extracted
from Fig.~6 of Ref.~\cite{Deans02}). Filled (open) circles are for WT
(Cx36-KO) mice, solid (dashed) lines show the results of the pair
approximation, thus $L\to\infty$, with $p=q=1$ ($p=q=0$). Upper curves
are for $n=10$, lower curves are for $n=3$. The dot-dashed line
corresponds to simulations with $n=10$, $p=q=1$ and $L=20$.}
\end{figure}

Stronger experimental support for our conjecture on the role of
electrical interactions is available for the mammalian retina. Deans
et al.~\cite{Deans02} have measured the firing rates of on-center
ganglion cells for varying light intensity (measured in isomerized
molecules of rhodopsin per rod per second, or Rh*/rod/s). The response
curves have been obtained for both wild type (WT) mice as well as mice
in which the expression of the protein connexin36 (responsible for the
gap junction intercellular channels) has been genetically knocked out
(Cx36-KO). The difference in the response curves can be seen in
Fig.~\ref{fig:deans}. They present the same qualitative behavior of
the curves shown in Fig.~\ref{fig:respostalinlog}, exhibiting an
increase in dynamic range in the presence of electrical coupling: 14dB
for Cx36-KO and 23dB for WT, values which are of the same order as
those of Fig.~\ref{fig:faixadinamicaL}. In particular, the exponent of
the ``coupled'' (WT) case is $\alpha\simeq 0.58$ (see inset), which is
slightly larger than what is obtained in the pair approximation. 

The quantitative agreement between the analytical and experimental
curves is limited. On the one hand, the theoretical $n=3$ curve can
provide a good fit of the Cx36-KO data for $p=q=0$, while the coupled
case $p=q=1$ does not adjust well to the WT data. For $n=10$ and
$p=q=1$, on the other hand, the WT data are well matched by
simulations with a finite $L=20$ system (staying below the
$L\to\infty$ pair approximation), but for $p=q=0$ the same $n=10$
automata are unable to give a good fit of the Cx36-KO data. The
difficulties of a quantitative match are not surprising: the retina is
organized in layers which have, to first order, a two-dimensional
structure, signal processing from the photoreceptors to the ganglion
cells involves a complex intermediate neuronal circuit (with bipolar,
horizontal and amacrine cells~\cite{Shepherd}), and individual neurons
themselves can have subtle dynamical properties (such as adaptation,
for instance). All these properties are clearly absent from our simple
one-dimensional CA model. Yet it correctly predicts the reduction in
the dynamic range of a neuronal system which loses electrical coupling
among its cells.

In order to have a quantitative agreement between experimental and
theoretical curves, additional modeling efforts are needed which
incorporate specific details of the system under
consideration. However, the response of simple models of excitable
media remains an important subject to be studied, precisely because
they have the potential to reveal simple mechanisms and scaling
relations~\cite{Ohta05} whose robustness can thereafter be subjected
to further testing in experiments and more detailed models. In this
context, the simple Greenberg-Hastings CA strikes an interesting
balance, on the one hand capturing essential features of collective
neuronal dynamics, while on the other hand lending itself to
analytical techniques borrowed from nonequilibrium statistical
mechanics.

\begin{acknowledgments}
The authors would like to thank O. Kinouchi, S. G. Coutinho,
A. C. Roque, R. F. Oliveira, R. Publio, M. J. de Oliveira and an
anonymous referee for useful discussions and comments. LSF is
supported by UFPE/CNPq/PIBIC. MC acknowledges support from Projeto
En\-xo\-val (UFPE), FACEPE, CNPq and special program PRONEX.
\end{acknowledgments}

\appendix*

\section{\label{ap:pair}The equations for two-site probabilities}

\subsection{Dynamics} 

In all derivations below, homogeneity and isotropy are assumed. The
sign ``$\approx$'' denotes that the equality holds in the pair
approximation [Eqs.~(\ref{eq:pair})].  We start by writing down the
equation for $P_{t}(0,0)$, which holds $\forall n\geq 3$:

\begin{eqnarray}
P_{t+1}(0,0) & = & P_{t}(n-1,n-1) \nonumber \\
&& + 2(1-\lambda)\left[P_{t}(n-1,0) - pP_{t}(1,0,n-1) \right] \nonumber \\
& & + (1-\lambda)^2 \left[ P_{t}(0,0) -  2pP_{t}(1,0,0) \right. \nonumber \\
&& \left. + p^2P_{t}(1,0,0,1) \right] \nonumber \\ 
\label{eq:p00t} \ & \approx & P_{t}(n-1,n-1)  \nonumber \\
&& + 2(1-\lambda)P_{t}(n-1,0)\left[1-p\frac{P_{t}(1,0)}{P_{t}(0)} \right]
  \nonumber \\
 & &   + (1-\lambda)^2P_{t}(0,0)\left[ 1-2p\frac{P_{t}(1,0)}{P_{t}(0)}
  \right. \nonumber \\
&& \left. +  p^2\frac{P_{t}(1,0)^2}{P_{t}(0)^2} \right] .
\end{eqnarray}
The dynamics for two-site probabilities in the refractory period obey
a simple recursive rule due to the deterministic evolution of the
automata:

\begin{equation}
\label{eq:pjk}
P_{t+1}(j,k) = P_{t}(j-1,k-1) \; , \; \; 2\leq j,k \leq n-1\; .
\end{equation}
On the one hand, diagonal terms $P_{t}(j,j)$ with $j \geq 2$
recursively depend on $P_{t}(1,1)$, whose dynamics can be written as
follows:

\begin{eqnarray}
P_{t+1}(1,1) & = & \lambda^2P_{t}(0,0) +
2p\lambda(1-\lambda)P_{t}(1,0,0) \nonumber \\
&& + p^2(1-\lambda)^2P_{t}(1,0,0,1)\nonumber \\ 
\label{eq:p11}\ & \approx & P_{t}(0,0) \left[\lambda^2 +
2p\lambda(1-\lambda)\frac{P_{t}(1,0)}{P_{t}(0)}  \right. \nonumber \\
&& + \left. p^2(1-\lambda)^2\frac{P_{t}(1,0)^2}{P_{t}(0)^2}\right]\; .
\end{eqnarray}
Off-diagonal terms, on the other hand, ultimately depend on
$P_{t}(j,1)$. For $j=2$, the equation is simply

\begin{eqnarray}
\label{eq:p21}
P_{t+1}(2,1) & = & (\lambda+p-p\lambda)P_{t}(1,0) \nonumber \\
&& + (1-\lambda)(q-p)P_{t}(1,0,1) \nonumber \\
&\approx & P_{t}(1,0)\left\{ \lambda +  (1-\lambda)\left[ p
\vphantom{\frac{P_{t}(1,0)}{P_{t}(0)}} \right. \right. \nonumber \\
&&  \left.\left.   
 + (q-p)\frac{P_{t}(1,0)}{P_{t}(0)} \right]  \right\}
\; ,
\end{eqnarray}
while for $j\geq 3$  one has

\begin{eqnarray}
\label{eq:pj1}
P_{t+1}(j,1) & = & \lambda P_{t}(j-1,0) + p(1-\lambda)P_{t}(j-1,0,1)\nonumber \\
\ & \approx &  P_{t}(j-1,0)\left[\lambda +  p(1-\lambda)\frac{P_{t}(1,0)}{P_{t}(0)} \right]\; .
\end{eqnarray}
Finally, one needs equations for $P_{t}(j,0)$, $j\geq 2$ [recall
Eq.~(\ref{eq:p10pair}) for $P_{t}(1,0)$]. Like in Eq.~(\ref{eq:p21}),
the case $j=2$ must be considered separately:
\begin{eqnarray}
\label{eq:p20}
P_{t+1}(2,0) & = & P_{t}(1,n-1) + (1-\lambda)(1-p)P_{t}(1,0) \nonumber \\
&& + (1-\lambda)(p-q)P_{t}(1,0,1) \nonumber \\ 
&\approx & P_{t}(1,n-1)+(1-\lambda)P_{t}(1,0)\left[(1-p) \vphantom{\frac{P_{t}(1,0)}{P_{t}(0)}}\right. \nonumber \\
&& \left. +(p-q)\frac{P_{t}(1,0)}{P_{t}(0)}
\right]\; .
\end{eqnarray}
For $j\geq 3$, on the other hand, one immediately obtains

\begin{eqnarray}
\label{eq:pj0}
P_{t+1}(j,0) & = & P_{t}(j-1,n-1)  \nonumber \\
&& + (1-\lambda)\left[P_{t}(j-1,0) - pP_{t}(j-1,0,1) \right] \nonumber \\
\ & \approx & P_{t}(j-1,n-1) \nonumber \\
&& + (1-\lambda)P_{t}(j-1,0)\left[1 -  p\frac{P_{t}(1,0)}{P_{t}(0)} \right]\; ,
\end{eqnarray}
which completes the set of all pair equations. Upon iteration of
Eqs.~(\ref{eq:refrac},\ref{eq:norm},\ref{eq:p1pair},\ref{eq:p10pair},\ref{eq:p00t}-\ref{eq:pj0}),
normalization conditions properly imposed in the initial conditions
are naturally preserved. To determine the response function
$P(1)=F(\lambda)$, we wait until the $(n^2+3n)/2$-dimensional map
reaches a stationary state for each value of $\lambda$. We describe
below how the analysis of the stationary state can be reduced to just
two equations [Eqs.~(\ref{eq:2}-\ref{eq:1})].

\subsection{Stationary state} 

We start by handling the case $n>4$. In the stationary state, the
first term on the r.h.s. of Eq.~(\ref{eq:pj0}) becomes, via recursive
iterations of Eq.~(\ref{eq:pjk}),
\begin{equation}
\label{eq:p1j}
P(j-1,n-1)  = P(1,1+n-j), \; \forall j\geq 3\; .
\end{equation}
The above result can on its turn be further developed by means of
Eq.~(\ref{eq:pj1}) as long as $1+n-j\geq 3$, rendering the stationary
state of Eq.~(\ref{eq:pj0}):
\begin{eqnarray}
P(j,0) &\approx& P(n-j,0)\left[\lambda+p(1-\lambda)\frac{P(1,0)}{P(0)}\right] \nonumber \\
&& +(1-\lambda)P(j-1,0)\left[1 -  p\frac{P(1,0)}{P(0)} \right]\; , \nonumber \\
&& 3\leq j \leq n-2\; .
\end{eqnarray}
Notice that we have a nonhomogeneous set of $n-4$ linear equations for
$x_j \equiv P(j,0)$: $x_j \approx ax_{n-j} + (1-a)x_{j-1}$, where
$a\equiv \lambda + p(1-\lambda)P(1,0)/P(0)$ and $x_2 = P(2,0)$ accounts
for the nonhomogeneity in the equations for $x_3$ and $x_{n-2}$. The
solution of these equations is simply $x_{n-2} \approx x_{n-1} \approx
\ldots \approx x_{3} \approx x_{2}$, as can be checked by
inspection. The combination of Eqs.~(\ref{eq:p20})~and~(\ref{eq:pj1})
in the stationary state, on the other hand, leads to 

\begin{eqnarray}
\label{eq:J}
P(j,0) & \approx & J\left[P(1,0),P(0)\right]  \nonumber \\
& \equiv &
P(1,0)\left[\frac{(1-p)P(0)+(p-q)P(1,0)}{P(0)-pP(1,0)}\right]\; , \nonumber \\
&& 2\leq j \leq n-2 \; . 
\end{eqnarray}
One therefore obtains

\begin{eqnarray}
\label{eq:xj}
P(n-2,0) & \approx & P(n-3,0) \approx \ldots \approx P(2,0) \nonumber \\
&\approx & J\left[P(1,0),P(0)\right]\; .
\end{eqnarray}
Finally, notice that $P(n-1,0)$ can be obtained by combination of
Eqs.~(\ref{eq:pj0}),~(\ref{eq:xj})~and~(\ref{eq:p21}):
\begin{equation}
\label{eq:pdelta}
P(n-1,0) \approx  P(1,0)\; ,
\end{equation}
which completes the proof for $n>4$. For $n=4$, it suffices to invoke
Eqs.~(\ref{eq:p20})~and~(\ref{eq:pj1}) to show that $P(2,0) \approx
J\left[P(1,0),P(0)\right]$.  With this result, Eq.~(\ref{eq:pdelta})
holds for $n\geq 4$. Finally, for $n=3$,
Eqs.~(\ref{eq:p20})~and~(\ref{eq:p21}) together also lead to
Eq.~(\ref{eq:pdelta}). 

Invoking the normalization condition
$P_{t}(0)=\sum_{j=0}^{n-1}P_{t}(j,0)$, one can deduce that, on the one
hand, 
\begin{equation}
\label{eq:p00} 
P(0,0) = P(0) - 2P(1,0) - (n-3)J\left[P(1,0),P(0)\right]\; .
\end{equation}
On the other hand, in the stationary state Eq.~(\ref{eq:p10pair})
depends linearly on $P(0,0)$, so it can be inverted, yielding [after
substitution of Eq.~(\ref{eq:pdelta})] $P(0,0)$ as a function of
$P(1,0)$ and $P(0)$. Equaling this function to Eq.~(\ref{eq:p00}),
$P(0,0)$ is eliminated and one obtains Eq.~(\ref{eq:2}).


\end{document}